\documentclass[a4paper,twocolumn,11pt]{article}

\usepackage[a4paper, margin=1in]{geometry}
\usepackage{amsmath,amssymb}
\usepackage{graphicx}
\usepackage{setspace}
\usepackage[sort&compress]{natbib}
\usepackage{hyperref}
\hypersetup{colorlinks=true, citecolor=blue}
\usepackage{authblk}
\usepackage{sectsty}
\sectionfont{\large}
\date{}

\title{Constraints on Rastall gravity from current observational data}

\author{Mahnaz Asghari \thanks{m.asghari@riaam.ac.ir}}
\author{Hooman Moradpour \thanks{h.moradpour@riaam.ac.ir}}
\affil{Research Institute for Astronomy and Astrophysics of Maragha (RIAAM), University of Maragheh, P. O. Box 55136-553, Maragheh, Iran}

\begin{document}
\maketitle

\begin{abstract}
We study the cosmological features of Rastall gravity, where the covariant energy-momentum conservation is modified to $\nabla_{\mu}T^{\mu}_{\nu}=\lambda \nabla_{\nu}R$ in curved spacetime. For this purpose, we obtain the modified field equations of Rastall model to linear order of perturbations, and then inspect the evolutionary behavior of cosmological observables, chiefly the matter power spectrum and the Hubble parameter, within the context of Rastall gravity. We also compare Rastall model with cosmological probes, namely cosmic microwave background, weak lensing, supernovae, baryon acoustic oscillations, and redshift-space distortions data. According to our numerical results, Rastall gravity prefers lower growth of structures compared to the $\Lambda$CDM model, which indicates consistency with low-redshift large-scale structure probes. Moreover, numerical analysis reveals that Rastall gravity with a cosmological constant as the dark energy component suffers from the Hubble tension, like the $\Lambda$CDM model.    	
\end{abstract}


\section{Introduction}
Recent improvements in observational cosmology propose the $\Lambda$CDM as the most successful model to describe the evolution of the universe. Specifically, the cosmic microwave background (CMB) anisotropies \cite{cmb1,cmb2,cmb3} as well as the type Ia supernovae distance measurements \cite{sn1,sn2}, confirm the robustness of the standard cosmological model in the framework of general relativity (GR). Although the cosmological observations affirm the viability of the standard $\Lambda$CDM model, there are some theoretical and observational insufficiencies that provide tendencies to probe beyond GR. In particular, the most recent direct determinations of the Hubble constant $H_0$ report approximately $6\sigma$ conflict with the CMB measurements based on $\Lambda$CDM \cite{h01,h02}. Additionally, there is a mild discrepancy regarding the matter clustering parameter $S_8$, where local observations predict less structure growth compared to the Planck data \cite{s81,s82,s83}. While the observed cosmological tensions can be due to systematic effects, it is also reasonable to consider new physics beyond the $\Lambda$CDM model. In this regard, there is a phenomenological modification of GR, known as Rastall gravity, in which the energy momentum tensor does not respect the usual conservation law \cite{rastall}. 

Rastall, in 1972, challenged the covariant energy-momentum conservation in curved spacetime, proposed by the fact that the conservation law $\nabla_{\mu}T^{\mu}_{\nu}=0$ has been mainly tested in the framework of special relativity and hence is considered to be valid in GR through the equivalence principle \cite{rastall}. Then, he derived modified field equations based on a simple generalized conservation law given by \cite{rastall} 
\begin{equation} \label{eq1}
\nabla_{\mu}T^{\mu}_{\nu}=\lambda \nabla_{\nu}R \,,
\end{equation}
where $\lambda$ is a constant and $R$ is the Ricci scalar. According to Eq. (\ref{eq1}) there is a non-minimal coupling between matter and geometry in Rastall model, where choosing $\lambda=0$ restores the GR formulation. 
In the following we review some related investigations on Rastall gravity; In 2011, Capone et al. \cite{Capone} studied the accelerated expansion in Rastall model using the available type Ia Supernovae and Gamma Ray Bursts data. Batista et al. \cite{Batista2012} in 2012, explored the Rastall cosmology with a two-fluid model. Then, in 2013, observational constraints on the two-fluid Rastall cosmological model were also considered by Batista et al. \cite{Batista2013}. Moreover, thermodynamics of Rastall gravity were studied in \cite{moradpour2016-I,Moradpour2016-II,Bamba2018}. On the other hand, \cite{Darabi2017} probed the relationship between Rastall gravity and GR, while the similarities between Rastall model and $f(R,T)$ theory were explored in \cite{Shabani2020,Fabris2020}. Some efforts focused on describing a Lagrangian for Rastall theory were also considered in \cite{DeMoraes2019}. Furthermore, regarding recent observational constraints on Rastall model, one can refer to \cite{Akarsu2020,Singh2024,Mohebi2025,Sadeghnezhad2025}. These studies provide the motivation for further investigating cosmological perturbations and the observational viability of Rastall gravity.

In the present study, we aim to investigate the background evolution and linear scalar perturbations of Rastall gravity, when dark energy is regarded as a cosmological constant. We also explore the evolution of cosmological observables in the framework of Rastall theory.    
We then use a comprehensive combination of observational datasets, including CMB, weak lensing, supernovae, baryon acoustic oscillations (BAO), and redshift-space distortions (RSD) measurements, to derive strong constraints on the cosmological parameters of Rastall model and also inspect the observational tensions in the context of Rastall gravity. 

The paper is organized as follows. In Sec. \ref{sec2} we describe the modified field equations in Rastall gravity. Sec. \ref{sec3} investigates the evolution of cosmological observables in Rastall model. The derived observational constraints on cosmological parameters are reported in Sec. \ref{sec4}. We summarize our conclusions in Sec. \ref{sec5}.   
\section{Rastall gravity field equations} \label{sec2}
In this part, we derive the modified field equations in Rastall model based on Eq. (\ref{eq1}) proposed by Rastall. Accordingly, we obtain the corresponding field equations given by \cite{rastall}
\begin{equation} \label{eq2}
R_{\mu \nu}+(\kappa\lambda-\frac{1}{2})Rg_{\mu \nu}=\kappa T_{\mu \nu} \,,
\end{equation}
where $\kappa$ is the gravitational constant in Rastall theory. The parameters $\kappa$ and $\lambda$ in the static, weak-field limit can be related as \cite{rastall}
\begin{equation} \label{eq3}
\kappa\bigg(\frac{3\kappa\lambda-\frac{1}{2}}{1-4\kappa\lambda}\bigg)=-\frac{1}{2}\kappa_e \,,
\end{equation}
where, $\kappa_e=8 \pi G$ is the Einstein gravitational constant. 
Moreover, the energy content of the universe is assumed as a prefect fluid with the energy-momentum tensor $T_{\mu\nu}=\big(\rho+p\big)u_{\mu}u_{\nu}+pg_{\mu \nu}$, where $\rho$, $p$, and $u_{\mu}$ stand for energy density, pressure, and four-velocity, respectively.

We consider a spatially homogeneous and isotropic universe at background level, describing by the spatially flat background Friedmann-Lema\^itre-Robertson-Walker (FLRW) metric given by
\begin{align} \label{eq4}
\mathrm{d}s^2=a^2(\tau)\big(-\mathrm{d}\tau^2+\mathrm{d}\vec{x}^2\big) \,.
\end{align} 
Then, by regarding the dimensionless constant $\beta=\kappa\lambda$ as the Rastall parameter, modified field equations at background level take the form
\begin{align} \label{eq5}
 H^2(1-4\beta)-2\beta\frac{H'}{a}=\bigg(\frac{4\beta-1}{6\beta-1}\bigg)\frac{1}{3}\kappa_e \sum_{i}\bar{\rho}_i \,,  
\end{align}
\begin{align} \label{eq6}
&3H^2(1-4\beta)+2(1-3\beta)\frac{H'}{a} \nonumber \\
&=-\bigg(\frac{4\beta-1}{6\beta-1}\bigg)\kappa_e \sum_{i}\bar{p}_i \,,
\end{align}
where the prime indicates derivative with respect to the conformal time, the bar stands for a quantity evaluated at background level, and index \textit{i} indicates the component \textit{i}th in the universe filled with radiation (R), baryons (B), dark matter (DM) and the cosmological constant ($\Lambda$).
Further, According to the Eqs. (\ref{eq5}) and (\ref{eq6}) the modified Friedmann equation is derived as
\begin{equation} \label{eq7}
H^2=\frac{\kappa_e}{6\beta-1}\left[\left(\beta-\frac{1}{3}\right)\sum_{i}\bar{\rho}_i+\beta \sum_{i}\bar{p}_i\right] \,,
\end{equation}
and consequently, the total density parameter defined as $\Omega_\mathrm{tot}=\sum_{i}\bar{\rho}_i/\rho_\mathrm{cr}$ (with $\rho_\mathrm{cr}=3H^2/\kappa_e$) in Rastall gravity becomes
\begin{align} \label{eq8}
\Omega_\mathrm{tot}=\frac{1}{3\beta-1}\bigg(6\beta-1-\beta\frac{1}{H^2}\kappa_e\sum_{i}\bar{p}_i\bigg) \,.
\end{align}
It can be easily seen that by choosing $\beta=0$ the standard field equations in GR will be recovered.
Moreover, the deceleration parameter $q$ in Rastall model take the form
\begin{align}
q&=-1-\frac{H'}{H^2 a} \nonumber \\
&=\frac{(6\beta-1)\sum_{i}\bar{\rho}_i+(6\beta-3)\sum_{i}\bar{p}_i}{(6\beta-2)\sum_{i}\bar{\rho}_i+6\beta\sum_{i}\bar{p}_i} \,.
\end{align}

Taking into account the importance of studying cosmological perturbations in Rastall gravity, we turn our attention to linear order scalar perturbations. Accordingly, the perturbed FLRW metric in synchronous gauge is given by
\begin{equation} \label{eq9}
\mathrm{d}s^2=a^2(\tau)\Big(-\mathrm{d}\tau^2+\big(\delta_{ij}+h_{ij}\big)\mathrm{d}x^i\mathrm{d}x^j\Big)
\end{equation}
where
\begin{align} \label{eq10}
h_{ij}(\vec{x},\tau)&=\int \mathrm{d}^3k\,e^{i\vec{k}.\vec{x}}
\bigg(\hat{k}_i\hat{k}_jh(\vec{k},\tau) \nonumber \\
&+\Big(\hat{k}_i\hat{k}_j-\frac{1}{3}\delta_{ij}\Big)6\eta(\vec{k},\tau)\bigg) \,,
\end{align}
in which $h$ and $\eta$ are scalar perturbations, and $\vec{k}=k\hat{k}$ \cite{pt}. In addition, the perturbed metric in conformal Newtonian gauge is
\begin{equation} \label{eq11}
\mathrm{d}s^2=a^2(\tau)\Big(-\big(1+2\Psi\big)\mathrm{d}\tau^2+\big(1-2\Phi\big)\mathrm{d}\vec{x}^2\Big) \,,
\end{equation}
where, $\Psi$ and $\Phi$ are gravitational potentials \cite{pt}. Therefor, modified field equations in synchronous gauge take the form
\begin{align} \label{eq12}
& \frac{a'}{a}h'-2k^2\eta+\beta(-h''-3\frac{a'}{a}h'+4k^2\eta) \nonumber \\
&=\bigg(\frac{4\beta-1}{6\beta-1}\bigg)\kappa_e a^2 \sum_{i}\delta \rho_{i(\mathrm{syn})} \,,  
\end{align}
\begin{align} \label{eq13}
& k^2\eta'=\bigg(\frac{4\beta-1}{6\beta-1}\bigg)\frac{1}{2}\kappa_e a^2  \sum_{i}\big(\bar{\rho}_i+\bar{p}_i\big)\theta_{i(\mathrm{syn})} \,,  
\end{align}
\begin{align} \label{eq14} 
&\frac{1}{2}h''+3\eta''+\big(h'+6\eta'\big)\frac{a'}{a}-k^2\eta \nonumber \\
&=-\bigg(\frac{4\beta-1}{6\beta-1}\bigg)\frac{3}{2}\kappa_e a^2 \sum_{i}\big(\bar{\rho}_i+\bar{p}_i\big)\sigma_{i(\mathrm{syn})} \,, 
\end{align}
\begin{align} \label{eq15}
&-2\frac{a'}{a}h'-h''+2k^2\eta+3\beta\bigg(h''+3\frac{a'}{a}h'-4k^2\eta\bigg) \nonumber \\
&=\bigg(\frac{4\beta-1}{6\beta-1}\bigg)3\kappa_e a^2 \sum_{i}\delta p_{i(\mathrm{syn})} \,,
\end{align}
and likewise, in conformal Newtonian gauge we find 
\begin{align} \label{eq16}
& k^2\Phi+3\frac{a'}{a}\Phi'+3\Big(\frac{a'}{a}\Big)^2\Psi \nonumber \\
&+\beta\Big(k^2\Psi-2k^2\Phi-3\Phi''-3\frac{a'}{a}(\Psi'+3\Phi') \nonumber \\
&-6\frac{a''}{a}\Psi\Big) 
=-\bigg(\frac{4\beta-1}{6\beta-1}\bigg)\frac{1}{2}\kappa_e a^2 \sum_{i}\delta \rho_{i(\mathrm{con})} \,, 
\end{align}
\begin{align} \label{eq17}
& k^2\Phi'+\frac{a'}{a}k^2\Psi \nonumber \\
&=\bigg(\frac{4\beta-1}{6\beta-1}\bigg)\frac{1}{2}\kappa_e a^2 \sum_{i}\big(\bar{\rho}_i+\bar{p}_i\big)\theta_{i(\mathrm{con})} \,,  
\end{align}
\begin{align} \label{eq18}
& k^2(\Phi-\Psi) \nonumber \\
&=\bigg(\frac{4\beta-1}{6\beta-1}\bigg)\frac{3}{2}\kappa_e a^2 \sum_{i}\big(\bar{\rho}_i+\bar{p}_i\big)\sigma_{i(\mathrm{con})} \,,  
\end{align}
\begin{align} \label{eq19}
& \bigg(2\frac{a''}{a}-\Big(\frac{a'}{a}\Big)^2\bigg)\Psi+\frac{a'}{a}\big(\Psi'+2\Phi'\big)+\Phi'' \nonumber \\
&+\frac{1}{3}k^2\big(\Phi-\Psi\big)+\beta\Big(k^2\Psi-2k^2\Phi-3\Phi'' \nonumber \\
&-3\frac{a'}{a}(\Psi'+3\Phi')-6\frac{a''}{a}\Psi\Big) \nonumber \\
&=\bigg(\frac{4\beta-1}{6\beta-1}\bigg)\frac{1}{2}\kappa_e a^2 \sum_{i}\delta p_{i(\mathrm{con})} \,,  
\end{align}
where $\theta$ is the divergence of velocity perturbations, and $\sigma$ is the shear stress. It is worth noting that the Rastall model will be investigated in the synchronous gauge in the following. 

Let us now bring our attention toward the conservation equations in Rastall model. In the interest of finding a relationship between the Ricci scalar $R$ and the trace of the energy-momentum tensor $T$ in Rastall gravity, we take the trace of Eq. (\ref{eq2}), which yields
\begin{equation} \label{eq20}
R=\frac{\kappa T}{4\kappa\lambda-1} \,,
\end{equation}
where $\kappa$ is given by Eq. (\ref{eq3}). Thus, the modified field equations can be rewritten as 
\begin{align} \label{eq21}
R_{\mu \nu}-\frac{1}{2}Rg_{\mu \nu}=\kappa\Big(T_{\mu \nu}-\frac{\kappa\lambda T}{4\kappa\lambda-1} g_{\mu \nu}\Big) \,, 
\end{align}
where introduces an effective energy-momentum tensor, and is also equivalent to the modified field Eqs. (\ref{eq2}).
Then, regarding the Bianchi identity $\nabla_{\mu}G^{\mu}_{\nu}=0$, the modified conservation equations take the form
\begin{equation} \label{eq22}
\nabla_{\mu}T^{\mu}_{\nu}=\frac{\kappa\lambda}{4\kappa\lambda-1} \nabla_{\nu}T \,,
\end{equation}
with $\kappa$ defined in Eq. (\ref{eq3}). In this direction, by considering $\bar{p}_i=w_i\bar{\rho}_i$ with a constant equation of state $w_i$, the continuity and Euler equations to linear order of perturbations become
\begin{equation} \label{eq23}
\bar{\rho}'_i=-\frac{3(1+w_i)(4\beta-1)}{3\beta(1+w_i)-1}\frac{a'}{a}\bar{\rho}_i \,,
\end{equation} 
\begin{align} \label{eq24}
\delta'_{i(\mathrm{syn})}&=\frac{4\beta-1}{1-3\beta(1+c^2_{si})} \nonumber \\
&\times\Bigg\{\delta_{i(\mathrm{syn})}\frac{a'}{a}\frac{1}{3\beta(1+w_i)-1}\bigg(3(w_i-c^2_{si}) \nonumber \\
&+\frac{9(c^2_{si}-c^2_{ai})\beta}{4\beta-1}\Big(\beta(1+c^2_{si})-c^2_{si}\Big)\bigg) \nonumber \\
&+(1+w_i)\theta_{i(\mathrm{syn})}\bigg[1+\frac{9(c^2_{si}-c^2_{ai})}{3\beta(1+w_i)-1}\frac{1}{k^2} \nonumber \\
&\times \bigg(\beta\frac{a''}{a}+\Big(\frac{a'}{a}\Big)^2\Big(-1-\beta \nonumber \\
&+3\beta(\beta-1)\frac{(c^2_{si}-c^2_{ai})}{3\beta(1+w_i)-1}\Big)\bigg)\bigg] \nonumber \\
&+\frac{1}{2}h'(1+w_i) \nonumber \\
&-\frac{9(1+w_i)(c^2_{si}-c^2_{ai})\beta}{3\beta(1+w_i)-1}\frac{a'}{a}\sigma_{i(\mathrm{syn})}\Bigg\} \,,  
\end{align}
\begin{align} \label{eq25}
\theta'_{i(\mathrm{syn})}&=\theta_{i(\mathrm{syn})}\frac{a'}{a}\bigg[\frac{3}{3\beta(1+w_i)-1} \nonumber \\
&\times\bigg((4\beta-1)(1+w_i)+(\beta-1)(c^2_{si}-c^2_{ai})\bigg) \nonumber \\
&-4\bigg]-k^2\sigma_{i(\mathrm{syn})}+\delta_{i(\mathrm{syn})}\frac{k^2}{(1+w_i)(4\beta-1)} \nonumber \\
&\times\Big(\beta(1+c^2_{si})-c^2_{si}\Big) \,. 
\end{align}
Once more, we perceive that the standard cosmology is restored for $\beta=0$.  

Contemplating modified equations in Rastall model, in the next section we explore the evolutionary behavior of cosmological observables, mainly the Hubble parameter and the CMB anisotropy and matter power spectra, in Rastall gravity with a cosmological constant as dark energy. Furthermore, in order to achieve robust constraints on cosmological parameters, we confront Rastall model with current observations.
\section{Numerical results} \label{sec3}
This section is devoted to modifying the Boltzmann code CLASS\footnote{Cosmic Linear Anisotropy Solving System} \cite{cl} in accordance with the field equations of Rastall model derived in Sec. \ref{sec2}. We use the Planck 2018 data \cite{cmb3} for the cosmological parameters, in view of studying the evolution of cosmological observables in Rastall theory.

\begin{figure*}[ht!]
	\includegraphics[width=8.2cm]{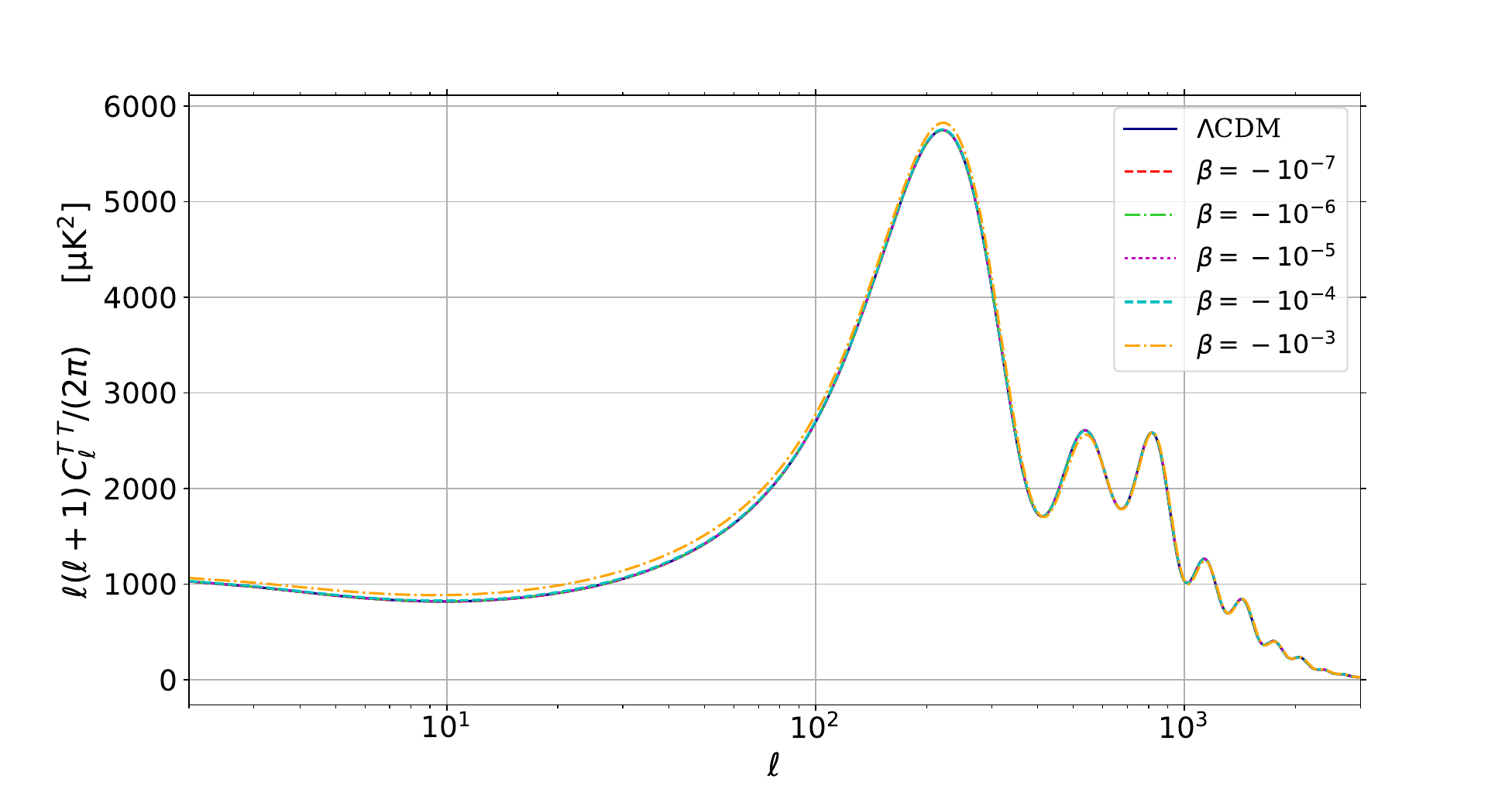}
	\includegraphics[width=8.2cm]{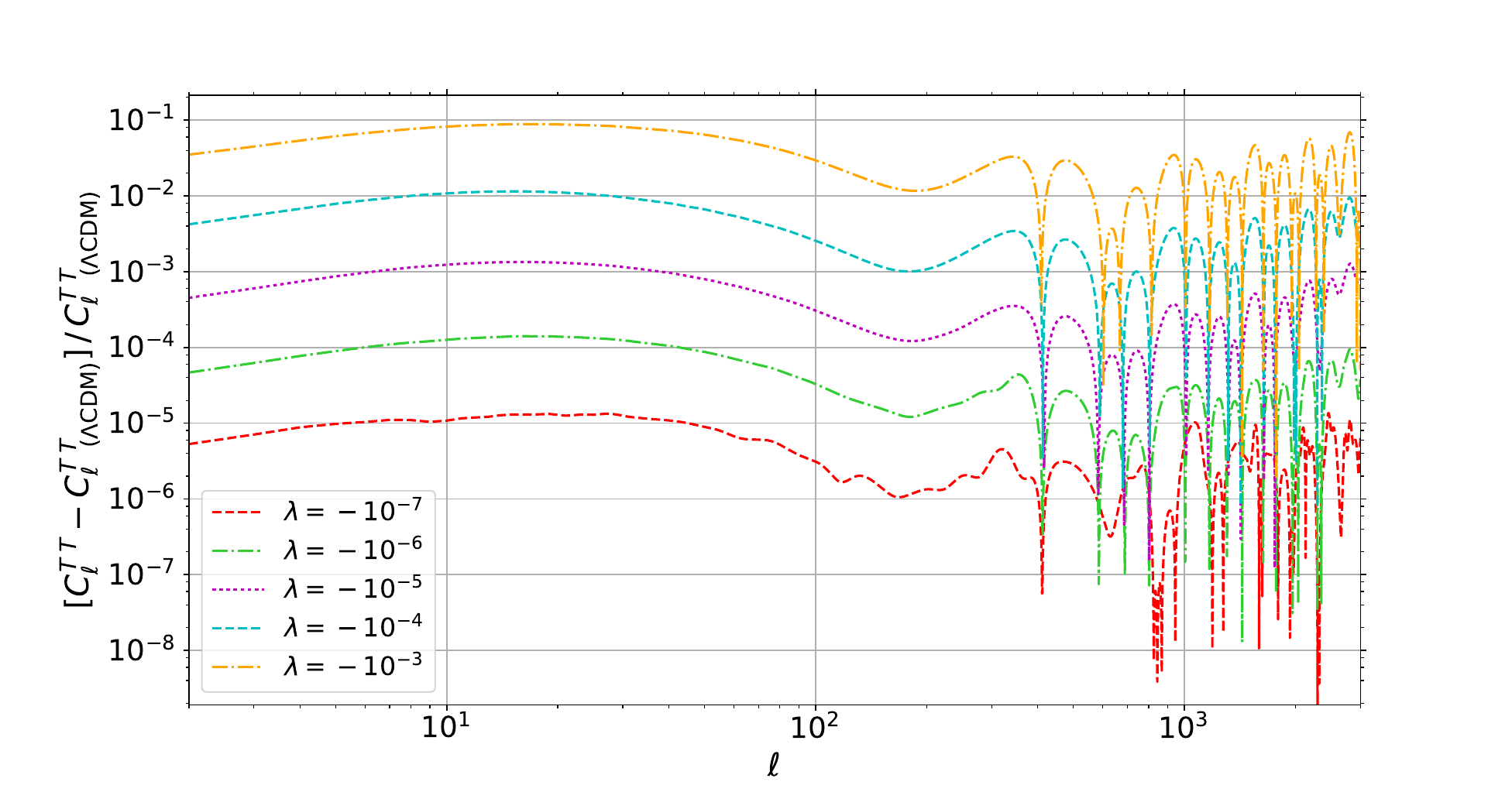}
	\includegraphics[width=8.2cm]{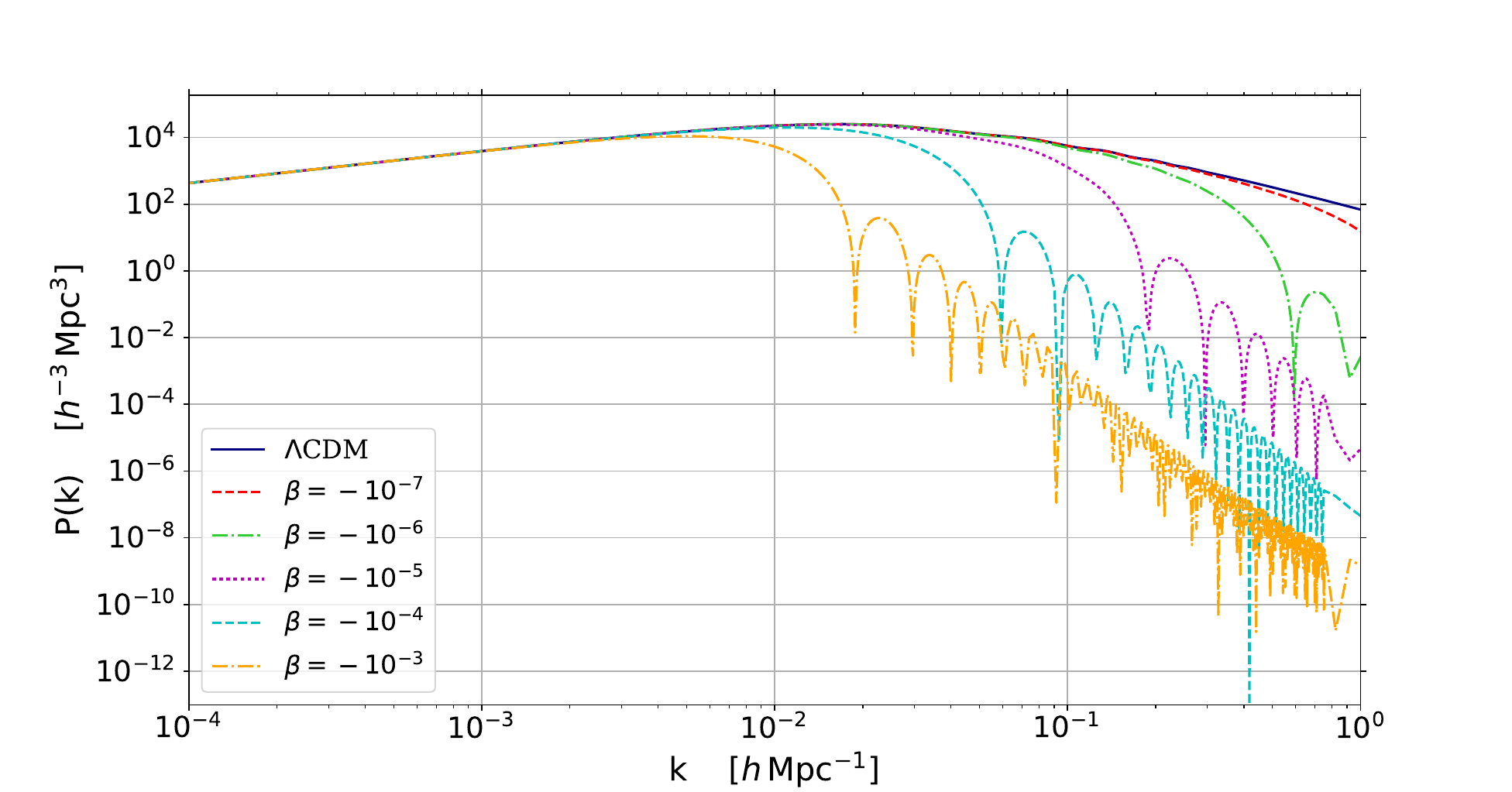}
	\includegraphics[width=8.2cm]{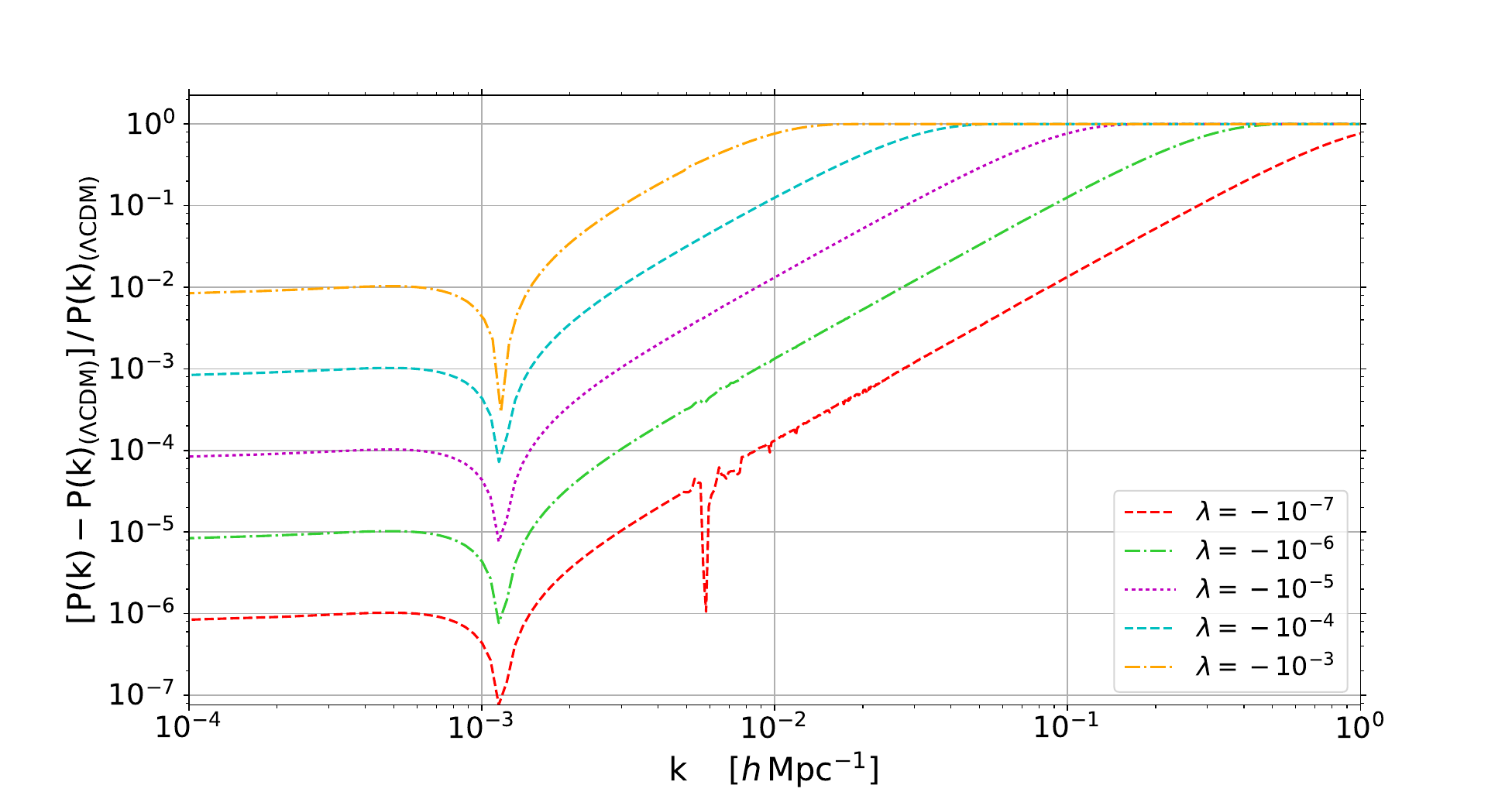}
	\caption{The CMB power spectra diagrams (upper left) and their relative ratio with respect to the $\Lambda$CDM (upper right) for different values of $\beta$. Lower panels show the analogous diagrams for the matter power spectra.}
	\label{f1}
\end{figure*}

Fig. \ref{f1} displays the CMB temperature anisotropy and matter power spectra for different values of $\beta$ in Rastall model. 
Regarding the matter power spectra diagrams, Rastall model predict a lower structure growth compared to the standard model of cosmology. This feature of Rastall gravity reveals consistency with some local measurements of structure formation \cite{s81,s82}. Moreover, the interaction between matter and curvature based on Eq. (\ref{eq1}), induces matter acoustic oscillations, which is apparent from matter power spectra diagrams. 

\begin{figure*}[ht!]
	\includegraphics[width=8.2cm]{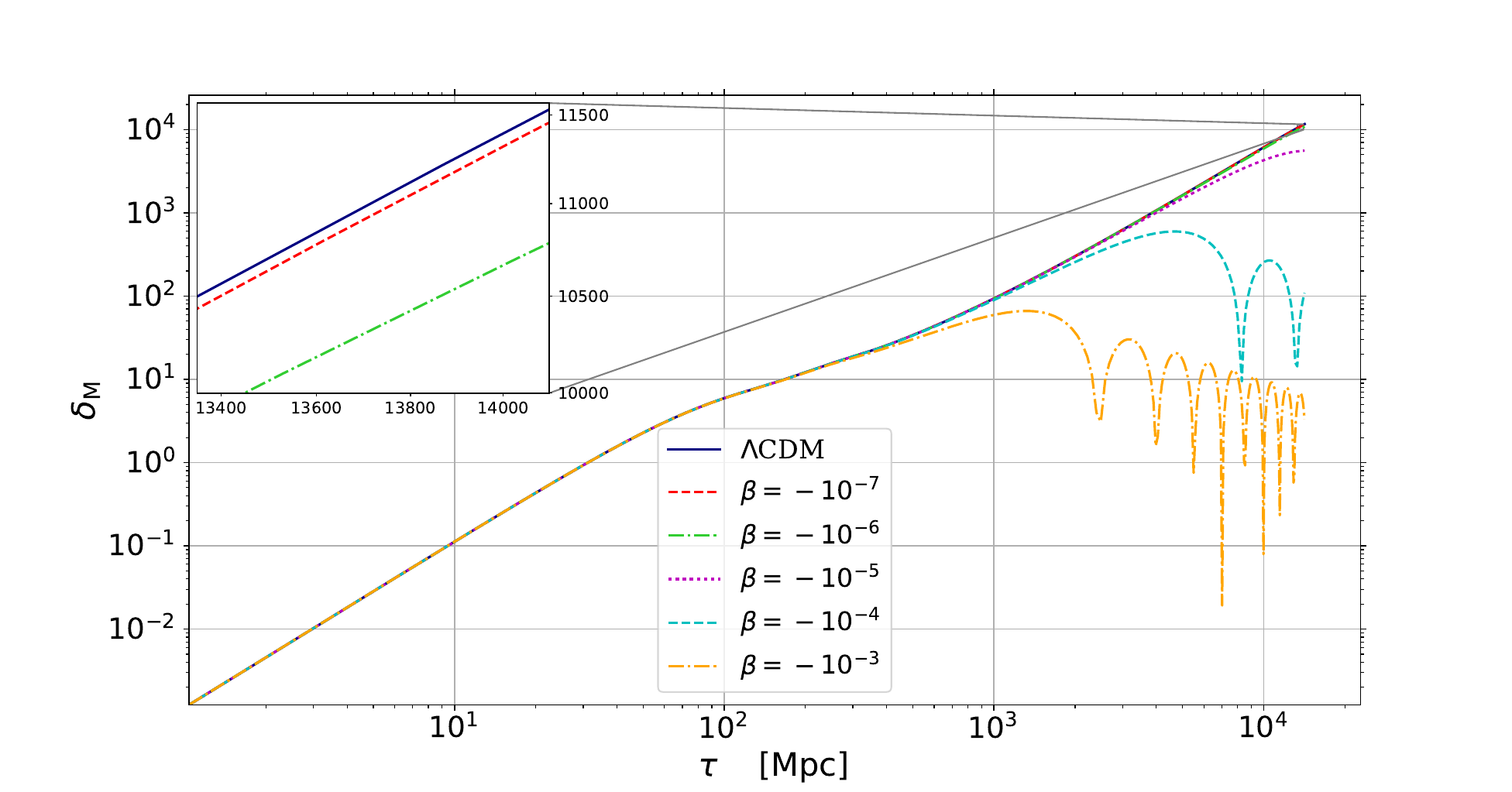}
	\includegraphics[width=8.2cm]{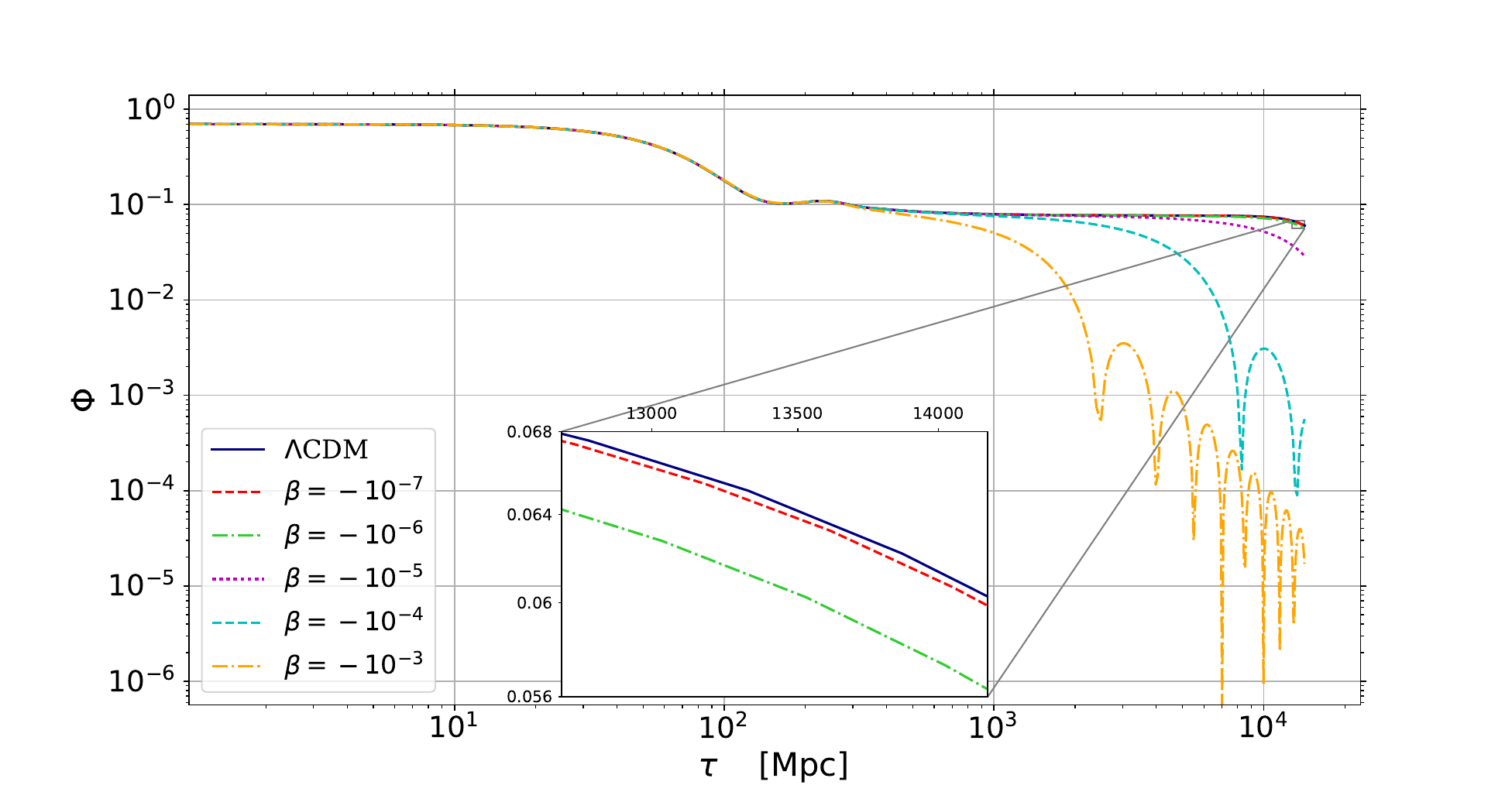}
	\caption{The matter density perturbations (left panel) and the Newtonian potential (right panel) in term of conformal time for different values of $\beta$, compared to $\Lambda$CDM.}
	\label{f2}
\end{figure*}

\begin{figure*}[ht!]
	\includegraphics[width=8.2cm]{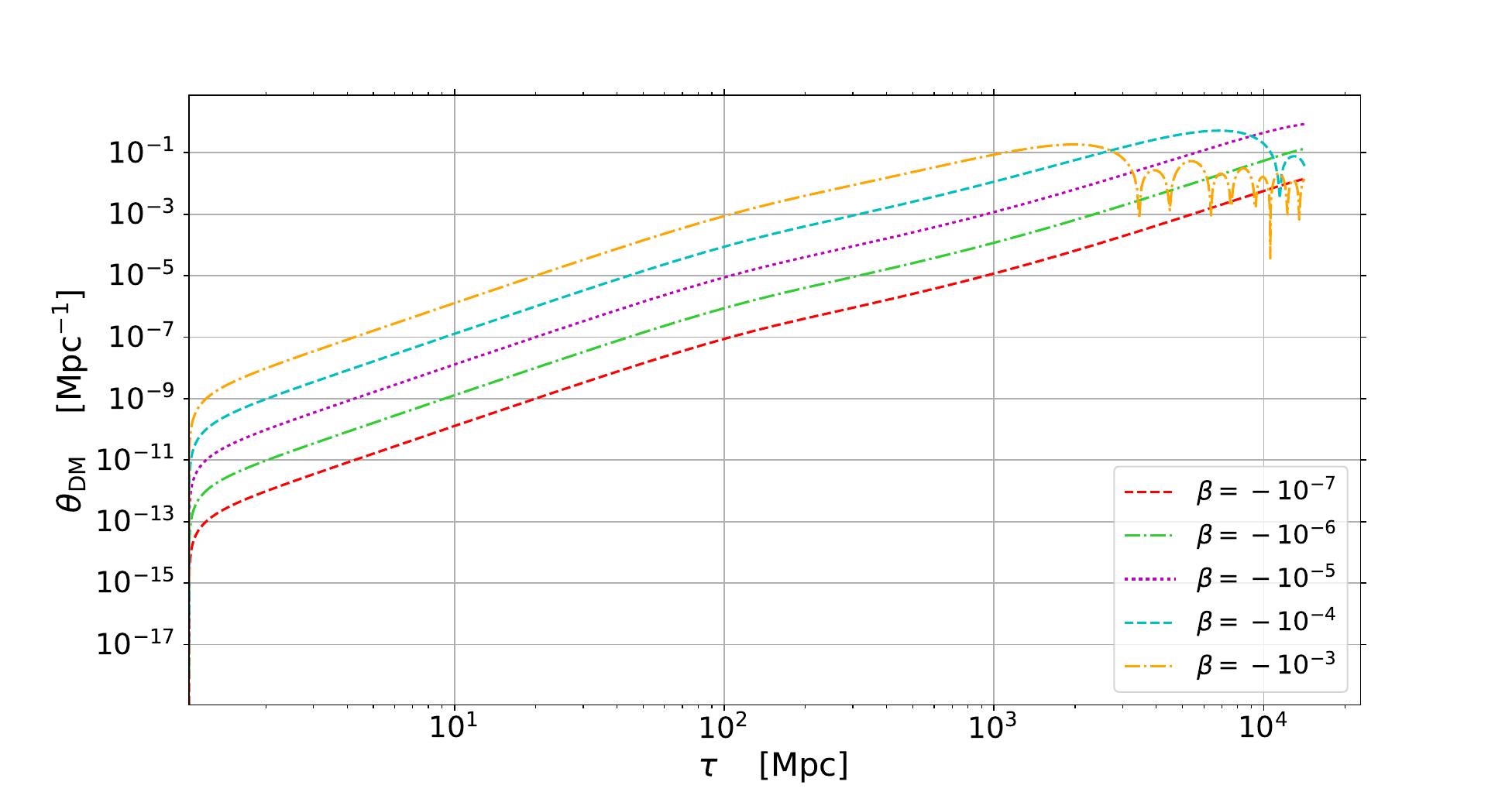}
	\includegraphics[width=8.2cm]{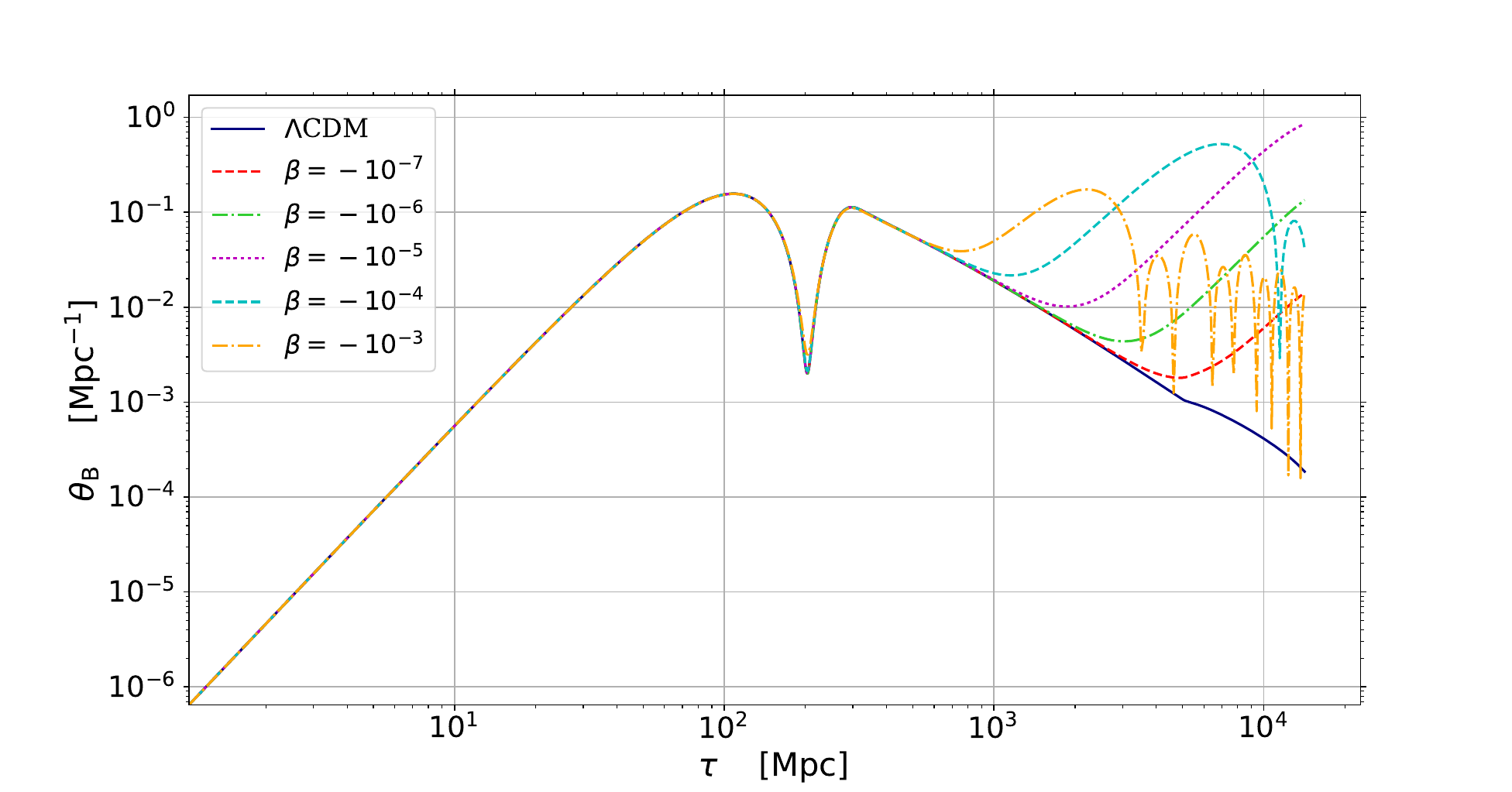}
	\caption{The velocity perturbations of dark matter (left panel) and baryons (right panel) in term of conformal time for different values of $\beta$.}
	\label{f3}
\end{figure*}

The evolution of matter density contrast and the Newtonian potential represented in Fig. \ref{f2}, also reflect the suppressed structure growth as well as the matter acoustic oscillations in Rastall gravity. 
Further, the velocity perturbations depicted in Fig. \ref{f3}, indicate the matter acoustic oscillations due to the matter-geometry coupling introduced in Eq. (\ref{eq1}).

It is also interesting that based on the interaction between matter and curvature in Rastall model, some researchers argue that the Rastall gravity may be regarded as a classical formulation of the quantum phenomenon for particle creation in cosmology \cite{Batista2012,thermo1}. Accordingly, the observed suppression in structure formation from our numerical analysis can show compatibility with the particle creation process. However, a comprehensive study from a thermodynamic viewpoint has not been performed for Rastall theory, and hence the oscillatory reduction in the structure growth can be mainly attributed to the matter-geometry coupling. 

\begin{figure*}[ht!]
	\includegraphics[width=8.2cm]{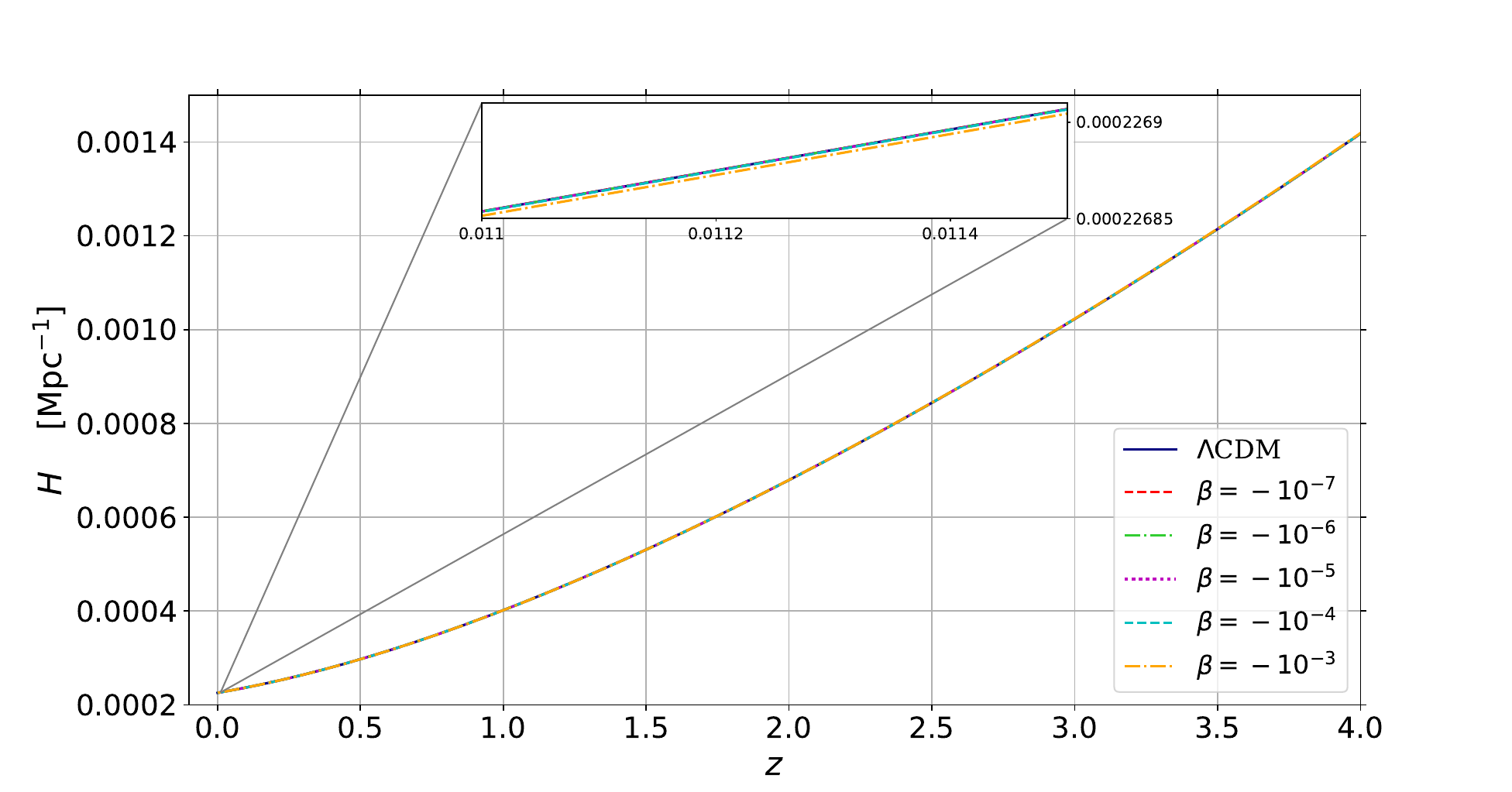}
	\includegraphics[width=8.2cm]{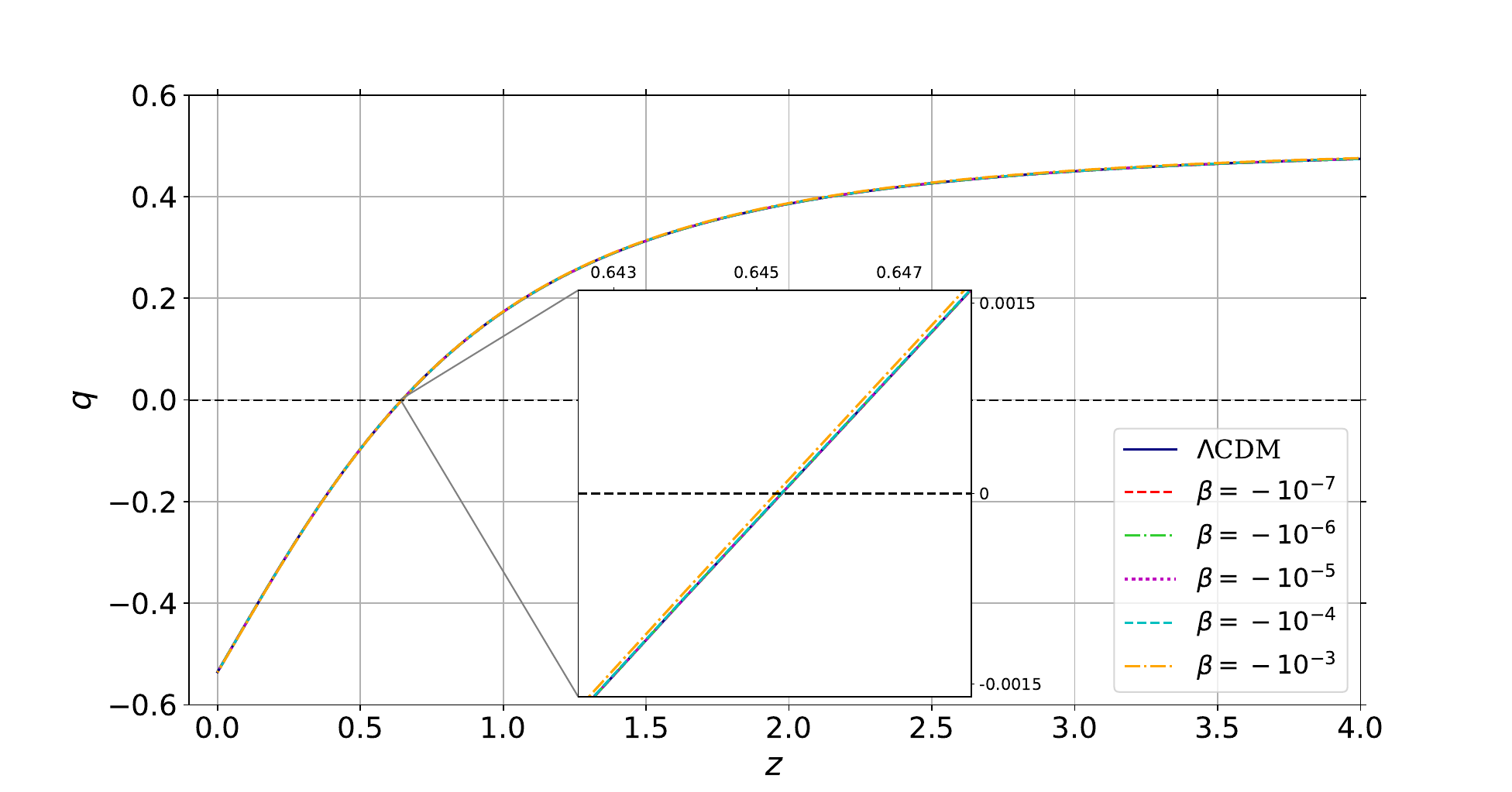}
	\caption{The Hubble parameter (left panel) and the deceleration parameter (right panel) in term of redshift for different values of $\beta$, compared to $\Lambda$CDM.}
	\label{f4}
\end{figure*}

The expansion history of the universe in Rastall gravity is investigated in Fig. \ref{f4}, where the evolution of the Hubble parameter $H$ and the deceleration parameter $q$ are represented in term of $z$. Accordingly, a suppression in the current value of Hubble parameter is reported for Rastall model. This result implies that the Hubble tension may become more serious in Rastall gravity with a cosmological constant as the dark energy candidate. Additionally, the $q$ diagrams indicate that the phase of accelerated expansion occurs in a lower redshift for Rastall model, which is consistent with $H(z)$ diagrams.  
  
On the other hand, from our numerical results it is worth noting that Rastall gravity shows some numerical similarities with the modified $f(R,T)$ theory \cite{frt1},as the cosmological observables display similar evolutionary behavior in both models. More specifically, comparing with Ref. \cite{frtobs}, it is notable that choosing appropriate values of dimensionless model parameters in two theories yields similar cosmological evolution. 
\section{Cosmological constraints} \label{sec4}
In pursuit of comparing the Rastall model with observational data, we employ the MCMC\footnote{Markov Chain Monte Carlo} package M\textsc{onte} P\textsc{ython} \cite{mp1,mp2}, interfaced with our modified CLASS code. The set of cosmological parameters considered in our MCMC approach includes \{$100\,\Omega_{\mathrm{B},0} h^2$,
$\Omega_{\mathrm{DM},0} h^2$, $100\,\theta_s$, $\ln (10^{10}A_s)$, $n_s$, $\tau_{\mathrm{reio}}$, $\beta$\}, containing the six cosmological parameters in $\Lambda$CDM model together with the Rastall parameter $\beta$. We also constrain five derived parameters through MCMC analysis, namely the reionization redshift $z_\mathrm{reio}$, the matter density parameter $\Omega_{\mathrm{M},0}$, the Hubble constant $H_0$, the structure growth parameter $\sigma_8$, and the matter clustering parameter $S_8$ defined as $S_8=\sigma_8\sqrt{\Omega_{\mathrm{M},0}/0.3}$. According to the primary numerical studies, the prior range [$-10^{-3}$, $10^{-11}$] is chosen for $\beta$. 

For the purpose of deriving constraints on cosmological parameters, we use the following dataset including the Planck likelihood with Planck 2018 data which contains high-$l$ TT,TE,EE, low-$l$ EE, low-$l$ TT, and lensing measurements \cite{cmb3}, the Planck-SZ likelihood for the Sunyaev-Zeldovich effect measured by Planck \cite{sz1,sz2}, the CFHTLenS likelihood with the weak lensing data \cite{lens1,lens2}, the PantheonPlus likelihood with the supernovae data \cite{panp},  the BAO likelihood with the baryon acoustic oscillations data \cite{bao1,bao2}, and the BAORSD likelihood for BAO and redshift-space distortions measurements \cite{rsd1,rsd2}. 

\begin{figure}
	\centering
	\includegraphics[width=8cm]{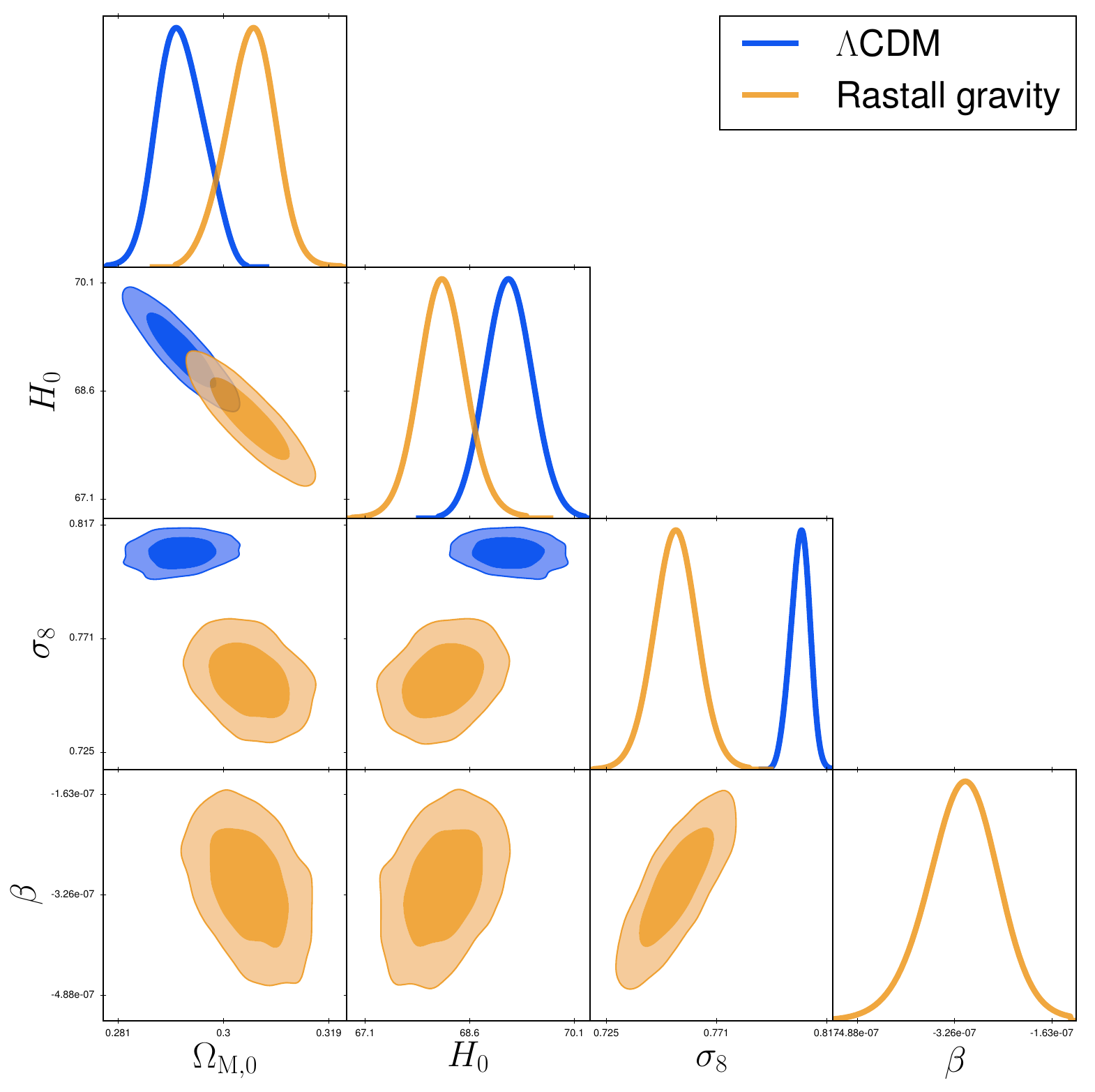}
	\caption{The one-dimensional posterior distribution and two-dimensional posterior contours with 68\% and 95\% confidence limits for some selected cosmological parameters of Rastall gravity compared to $\Lambda$CDM model.}
	\label{f5}
\end{figure}

The obtained constraints on cosmological parameters from the combined dataset "Planck + Planck-SZ + CFHTLenS + PantheonPlus + BAO + BAORSD" are summarized in table \ref{t1}.  The corresponding marginalized $1\sigma$ and $2\sigma$ posterior distributions for the selected cosmological parameters of Rastall model are shown in Fig. \ref{f5}.   
\begin{table}
	\centering
	\caption{\small Best fit values along with 68\% and 95\% confidence level intervals of cosmological parameters from "Planck + Planck-SZ + CFHTLenS + PantheonPlus + BAO + BAORSD" dataset for $\Lambda$CDM model and Rastall gravity.}
	\scalebox{.55}{
		\begin{tabular}{|c|c|c|c|c|}
			\hline
			& \multicolumn{2}{|c|}{} & \multicolumn{2}{|c|}{} \\
			& \multicolumn{2}{|c|}{$\Lambda$CDM} & \multicolumn{2}{|c|}{Rastall gravity} \\
			\cline{2-5}
			& & & & \\
			{parameter} & best fit & 68\% \& 95\% limits & best fit & 68\% \& 95\% limits \\ \hline
			& & & & \\
			$100\,\Omega_{\mathrm{B},0} h^2$ & $2.265$ & $2.258^{+0.012+0.025}_{-0.013-0.026}$ & $2.248$ & $2.242^{+0.013+0.028}_{-0.013-0.026}$ \\
			& & & & \\
			$\Omega_{\mathrm{DM},0} h^2$ & $0.1169$ & $0.1172^{+0.00070+0.0014}_{-0.00070-0.0015}$ & $0.1192$ & $0.1194^{+0.00075+0.0015}_{-0.00078-0.0015}$ \\
			& & & & \\
			$100\,\theta_s$ & $1.042$ & $1.042^{+0.00027+0.00061}_{-0.00031-0.00058}$ & $1.042$ & $1.042^{+0.00029+0.00060}_{-0.00030-0.00055}$ \\
			& & & & \\
			$\ln (10^{10} A_s)$ & $3.024$ & $3.022^{+0.0088+0.020}_{-0.012-0.019}$ & $3.048$ & $3.049^{+0.014+0.028}_{-0.015-0.029}$ \\
			& & & & \\
			$n_s$ & $0.9720$ & $0.9707^{+0.0035+0.0073}_{-0.0036-0.0073}$ & $0.9683$ & $0.9672^{+0.0033+0.0071}_{-0.0038-0.0070}$ \\
			& & & & \\
			$\tau_\mathrm{reio}$ & $0.04621$ & $0.04739^{+0.0022+0.0091}_{-0.0073-0.0074}$ & $0.05858$ & $0.05716^{+0.0063+0.014}_{-0.0079-0.014}$ \\
			& & & & \\
			$\beta$ & --- & --- & $-3.313\mathrm{e}{-7}$ & $-3.121\mathrm{e}{-7}^{+6.6\mathrm{e}{-8}+1.2\mathrm{e}{-7}}_{-5.6\mathrm{e}{-8}-1.3\mathrm{e}{-7}}$ \\
			& & & & \\
			$z_\mathrm{reio}$ & $6.741$ & $6.875^{+0.32+0.93}_{-0.69-0.82}$ & $8.076$ & $7.938^{+0.69+1.4}_{-0.72-1.3}$ \\
			& & & & \\
			$\Omega_{\mathrm{M},0}$ & $0.2903$ & $0.2925^{+0.0038+0.0084}_{-0.0042-0.0081}$ & $0.3039$ & $0.3047^{+0.0044+0.0091}_{-0.0047-0.0090}$ \\
			& & & & \\
			$H_0\;[\mathrm{\frac{km}{s\,Mpc}}]$ & $69.32$ & $69.15^{+0.33+0.67}_{-0.34-0.64}$ & $68.28$ & $68.22^{+0.35+0.71}_{-0.37-0.71}$ \\
			& & & & \\
			$\sigma_8$ & $0.8057$ & $0.8060^{+0.0039+0.0082}_{-0.0044-0.0077}$ & $0.7494$ & $0.7544^{+0.0094+0.020}_{-0.0098-0.019}$ \\
			& & & & \\
			$S_8$ & $0.7926$ & $0.7958^{+0.0068+0.015}_{-0.0077-0.015}$ & $0.7543$ & $0.7603^{+0.0087+0.019}_{-0.010-0.018}$ \\
			& & & & \\
			\hline
		\end{tabular}
	}
	\label{t1}
\end{table} 
Regarding derived constraints on the model parameters, we observe that Rastall gravity prefers a lower value for the structure growth parameter $\sigma_8$ with respect to standard model of cosmology, which is more compatible with some low-redshift structure growth observations \cite{s81,s82}. On the other hand, the derived constraints on $H_0$ parameter indicate that the Hubble tension becomes more severe in Rastall model when dark energy is considered to be a cosmological constant. Moreover, according to the constraints on the model parameter $\beta$, Rastall gravity signifies a deviation from the standard $\Lambda$CDM model at more than $3\sigma$.

It is worth mentioning that the obtained constraint on the Rastall parameter $\beta$ reported in table \ref{t1} is compatible with the thermodynamic interpretation of Rastall model discussed in Ref. \cite{Moradpour2016-II}, where the positivity of entropy yields the condition $\beta<1/6$ or $\beta>1/4$.  

It is also possible to evaluate the age of the universe from the CLASS code, using the obtained best fit values of cosmological parameters based on the MCMC analysis. Accordingly, we find a universe age of $t_0=13.75$ Gyr for Rastall gravity, and $t_0=13.72$ Gyr for $\Lambda$CDM model. Consequently, we notice a slightly larger universe age in the framework of Rastall model, which might be in better agreement with the ages of oldest astrophysical objects.   

For the last point, we utilize the Akaike information criterion (AIC) to assess which cosmological model provides the better fit to observations. The AIC is given by \cite{aic1,aic2}
\begin{equation}
\mathrm{AIC}=-2\ln{\mathcal{L}_{\mathrm{max}}}+2K ,
\end{equation}
in which $\mathcal{L}_{\mathrm{max}}$ is the maximum likelihood function and $K$ is the number of free parameters. Accordingly, numerical results report $\mathrm{AIC_{(\Lambda CDM)}}=4246.86$, and $\mathrm{AIC_{(Rastall)}}=4218.26$, which yields
$\mathrm{\Delta AIC}=28.6$. Thus, the AIC results show that the Rastall model is strongly preferred over the standard cosmological model. Therefore, it is suggested that Rastall model merits further investigations based on more precise and reliable observational datasets. 
\section{Conclusions} \label{sec5}
The covariant conservation of energy-momentum is considered as a fundamental principle of GR. However, the assumption $\nabla_{\mu}T^{\mu}_{\nu}=0$ is primarily tested in the context of special relativity, and thus inspires modifications of the covariant conservation law in curved spacetime \cite{rastall}. Correspondingly, Rastall proposed the modified conservation law $\nabla_{\mu}T^{\mu}_{\nu}=\lambda \nabla_{\nu}R$ in curved spacetime \cite{rastall}, suggesting a non-minimal matter-geometry coupling. 

In this paper, we study Rastall gravity in background and perturbation levels, employing our modified version of the CLASS code according to the Rastall field equations described in Sec. \ref{sec2}. We investigate the evolutionary behavior of cosmological observables, mainly the Hubble parameter and matter power spectrum, in the framework of Rastall gravity. According to the matter power spectra diagrams in Fig. \ref{f1}, we detect a suppression in the growth of structures in Rastall model which is consistent with some low-redshift large-scale structure observations \cite{s81,s82}. Moreover, the matter density perturbations and Newtonian potential diagrams depicted in Fig. \ref{f2} confirm the structure growth suppression in Rastall theory. It is also notable that the coupling between matter and geometry as a result of modified covariant conservation Eq. (\ref{eq1}) induces matter acoustic oscillations detected in Figs. \ref{f1} and \ref{f2}. These oscillations are also evident in velocity perturbations diagrams displayed in Fig. \ref{f3}. On the other hand, the evolution of Hubble parameter shown in Fig. \ref{f4} indicates a suppression in the $H_0$ value, which is in tension with low-redshift determinations of Hubble constant \cite{h01}.  

Further, we use current observational data to put strong constraints on cosmological parameters of Rastall model. To this end, we apply an MCMC analysis through the M\textsc{onte} P\textsc{ython} code, along with employing the combined "Planck + Planck-SZ + CFHTLenS + PantheonPlus + BAO + BAORSD" dataset. Numerical results reveal that Rastall gravity predict a lower structure growth compared to the $\Lambda$CDM model, showing consistency with some low-redshift cosmological probes. In addition, obtained constraints on the Hubble constant indicate that the $H_0$ tension becomes more severe in Rastall model with a cosmological constant as dark energy. 

Finally, we point to numerical similarities between Rastall theory and $f(R,T)$ gravity \cite{frtobs}, which are evident from the similar evolutionary behavior of cosmological observables together with the obtained constraints on cosmological parameters. It is interesting that Rastall gravity and modified $f(R,T)$ theory which are based on different physical principles and theoretical features, exhibit similar numerical behavior.     
\section*{Acknowledgments}
This research is supported by the research grant of the University of Maragheh. 
\section*{Competing interests}
The authors declare that they have no known competing financial interests or personal relationships that could have appeared to influence the work reported in this paper.
\section*{Data availability}
No new data were generated or analysed during the current study. 

\bibliographystyle{unsrt} 
\bibliography{1}

\end{document}